# Temperature dependence of electronic and magnetic properties of (DOEO)$_4$[HgBr$_4$]·TCE single crystals

Alisa Chernenkaya[1,2,a,*], Aleksandr Kotov[3], Oksana Koplak[3], Katerina Medjanik[2], Roman Morgunov[3], Eduard Yagubskii[3], Hans-Joachim Elmers[2] and Gerd Schönhense[2,b]

[1] Graduate School Materials Science in Mainz, 55128, Mainz, Germany

[2] Institut für Physik, Johannes Gutenberg-Universität, 55128, Mainz, Germany

[3] Institute of Problems of Chemical Physics, RAS, 142432 Chernogolovka, Russia

[a]chernenk@uni-mainz.de, [b]schoenhe@uni-mainz.de

**Abstract.** The temperature dependence of electronic and magnetic properties of the organic charge-transfer salt (DOEO)$_4$[HgBr$_4$]·TCE was investigated using magnetometry. Electronic transport properties revealed three distinct phases which are related to different magnetic coupling phenomena. In the low-temperature insulating phase ($T$<70 K) the antiferromagnetic coupling between two distinct sites of magnetic moments causes antiferromagnetic order below the Néel temperature $T_N$=40 K. In the temperature region 70-120 K the (DOEO)$_4$[HgBr$_4$]·TCE shows metallic-like behavior and with further increasing of temperature it becomes a "bad" metal due to loss of itinerant character and increase of hopping conductivity of charge carriers.

**Introduction**

Charge transfer complexes based on the organic single-electron donor bis(ethylenedithio)tetrathiafulvalene (BEDT-TTF, ET in short) represent the main class of layered organic conductive compounds [1-4]. They are characterized by two- or one-dimensional electronic properties. The narrow electronic bands near the Fermi energy cause strong electronic correlation effects leading to competing ordering instabilities. Consequently, one finds organic semiconductors, metals and superconductors in the wide class of these materials [5-7]. The easy tunability of the correlation effects by external fields like magnetic and electric fields as well as pressure has evoked a lot of scientific interest. In particular, $\kappa$-(ET)$_2$X salts show tunable Mott-insulating, quantum-liquid and superconducting properties [8-10]. Magnetic properties are usually caused by electronic states localized at the metal ions, especially if they are 3d transition metals. Charge-transfer salts where the correlation effects and the magnetic properties are caused by the identical electronic states are therefore of high interest.

The new cation-radical salt (DOEO)$_4$[HgBr$_4$]·TCE (where DOEO is 1,4-(dioxandiil-2,3-dithio)ethylenedithiotetrathiafulvalene and TCE is 1,1,2-trichloroethane) was recently synthetized [11]. The molecular structure of DOEO has a base skeleton similar to BEDT-TTF but the non-coplanar dioxane fragment makes adjustments to the packaging layers of cation-radicals (fig. 1) and therefore shows some notable differences to the familiar kappa-ET phases. The metal anion has a closed shell configuration and carries no spin. The optical and conductive properties of the charge transfer salt (DOEO)$_4$[HgBr$_4$]·TCE were described in Ref. 11-13. According to X-ray diffraction data at room temperature [11], the following types of structural disorder were observed: the cationic subsystem of ethylene groups in DOEO can occupy two nonequivalent positions, solvent molecules and anions are also partially disordered. At 30 K a partial ordering of each of these subsystems was found. It was reported [11, 12] to be a possible reason of the non-monotonous temperature dependence of the (DOEO)$_4$[HgBr$_4$]·TCE resistivity (fig. 2 a). Figure 2 b shows two types of localized spins with different local magnetic fields below 70 K in charge carriers in $(DOEO)_4[HgBr_4]\cdot TCE$ [14, 15] according to electron spin resonance (ESR) data.

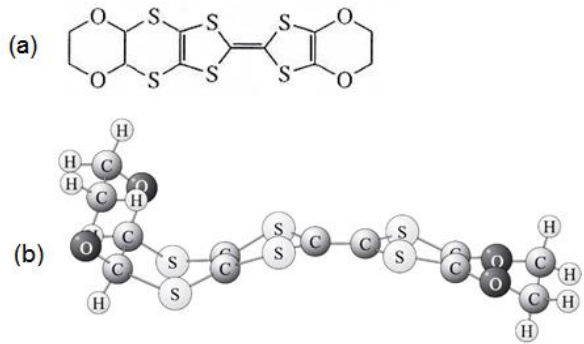

Fig. 1. (a) Chemical structure of the DOEO molecule; (b) Spatial arrangement of atoms in the DOEO molecule [11].

The article provides a new understanding of the antiferromagnetic coupling between the spins of the organic radicals in $(DOEO)_4[HgBr_4]\cdot TCE$ single crystals, which is different from the behavior of the related ET compounds. For the new charge-transfer salt $(DOEO)_4[HgBr_4]\cdot TCE$ there exists no systematic characterization of any phases and transitions up to now and it is still unknown if this new compound fits to the conceptual phase diagram of the ET-family. The combination of several physical properties including electrical conductivity, static and dynamic magnetic susceptibility allows the proposition of a phase model for $(DOEO)_4[HgBr_4]\cdot TCE$ as a representative of a new class of DOEO charge transfer salts with asymmetric organic cations.

### Experimental

The $(DOEO)_4[HgBr_4]\cdot TCE$ single crystals were prepared with the electro crystallization method in constant current mode as previously reported [11]. Their crystallographic structure is described in detail in Ref. 11. The temperature dependence of the magnetic moment $M$ of the samples was measured in the 2-300 K range at the constant field $H=0.1$ T using SQUID magnetometer (type MPMS 5XL, Quantum Design). The diamagnetic contribution of the sample holder at $T=300$ K was less than 1% of the signal. The magnetic moments at each temperature $M(T)$ were referenced to the dc molar magnetic susceptibility $\chi_{exp}=M/(vH)$ (fig. 2 c).

### Results and discussion

As two critical temperatures ($T = 70$ K – onset of charge carrier localization and $T = 120$ K – maximum of the resistivity curve) were previously observed, it is reasonable to expect changes in electronic structure and nonlinear temperature dependence of the magnetic susceptibility when crossing these temperatures.

As the SQUID technique is very sensitive and detects the sum of all magnetic moments, firstly the contribution of magnetic impurities was determined. These can be dislocations in the crystal and defects in the molecular packing in the $(DOEO)_4[HgBr_4]\cdot TCE$ crystal. The fit of the local impurity magnetization was described by the Curie temperature dependence $\chi_{imp}(T) = C/T$, assuming that the impurity spins are uncoupled. A second, temperature-independent contribution to the experimental susceptibility is given by the diamagnetic component $\chi_{dia}$, which is mainly caused by the $\pi$-states of the aromatic rings. $\chi_{dia}$ was estimated according to Ref. 16 from the molecular weight resulting in the constant value $\chi_{dia}=-1\cdot 10^{-3}$ emu·mol$^{-1}$Oe$^{-1}$, i.e. the dotted line at the bottom of Fig. 2 c. We subtracted $\chi_{imp}(T)$ and $\chi_{dia}(T)$ from the experimentally determined susceptibility (see Fig. 2 c), $\chi_s=\chi_{exp}-\chi_{dia}-\chi_{imp}$. At 300 K obtained value of the magnetic susceptibility $\chi_s$ agrees well with the assumption of one unpaired spin $S=\frac{1}{2}$ per DOEO dimer. According to the crystal structure of $(DOEO)_4[HgBr_4]\cdot TCE$ [11] and taking into account the unit cell volume and number of dimers in one unit cell one can calculate the expected saturation value of the magnetic moment per unit cell $\mu_{u.c.}=N_s g S \mu_B/V_{u.c.}=1.1\cdot 10^4$ emu·mol$^{-1}$. In this formula $N_s=8$ is the number of spins in one unit cell, $g$ is the gyromagnetic factor, $S=\frac{1}{2}$ is the spin value, $\mu_B$ the Bohr magneton and $V_{u.c.}$ the unit cell volume. Then, in the approximation of independent spins $S=\frac{1}{2}$ at room temperature we get a magnetization $M=\mu_{u.c.}\mu_B H/(k_B T)=2.5$ emu·mol$^{-1}$, where $k_B$ is the Boltzmann constant and $H=1000$ Oe is the field.

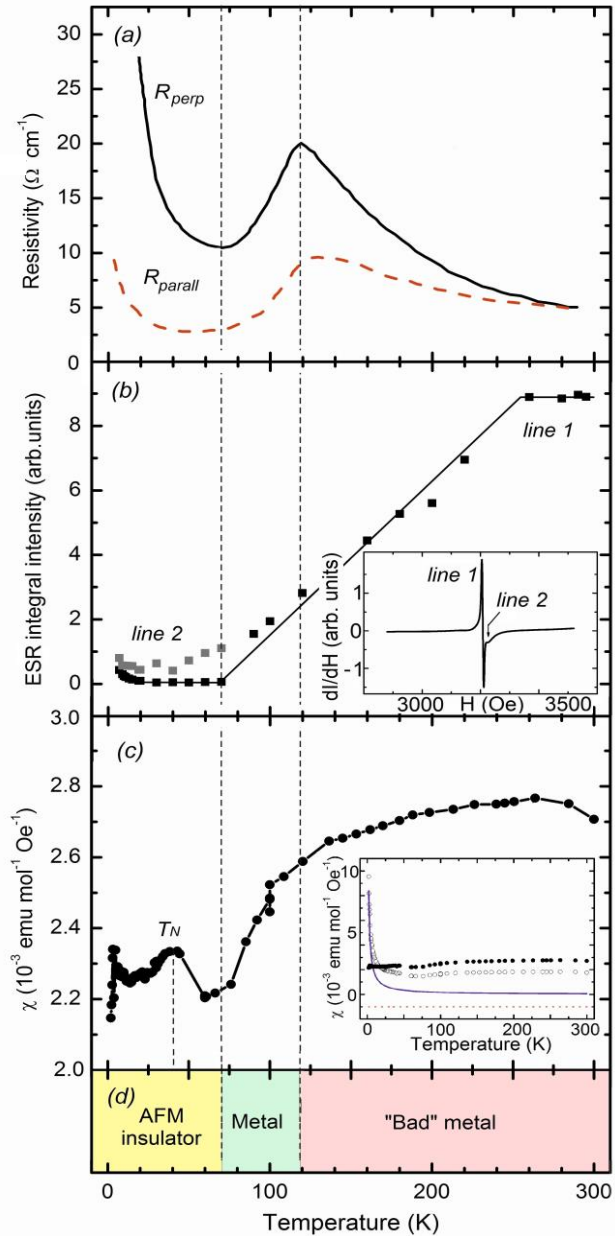

Fig. 2. (a) Electrical resistivity in the DOEO plane ($R_{parall}$) and in the direction perpendicular to this plane ($R_{perp}$) [11]. (b) Integral intensities of lines *1* (black squares) and *2* (gray squares) of a $(DOEO)_4[HgBr_4]\cdot TCE$ from the ESR spectra [14]. The inset shows a typical ESR spectrum taken at $T$ = 20 K that consists of the sum of lines 1 and 2 [14]. (c) Dynamic magnetic susceptibility $\chi$ of a $(DOEO)_4[HgBr_4]\cdot TCE$ from the SQUID data in a constant magnetic field of 0.1 T. The inset shows experimental data of the measured magnetic susceptibility (open circles) that was decomposed into the diamagnetic component (dotted line), Curie-Weiss component (solid line) and the true $(DOEO)_4[HgBr_4]\cdot TCE$ component (full circles). (d) Proposed phase diagram of $(DOEO)_4[HgBr_4]\cdot TCE$; vertical lines denote phase transition or crossover temperatures 70 K and 120 K and Néel temperature $T_N$ = 40 K.

This corresponds $\chi_{theor}=2.5\cdot10^{-3}$ emu·mol$^{-1}$·Oe$^{-1}$. The experimental value $\chi=2.68\cdot10^{-3}$ emu·mol$^{-1}$·Oe$^{-1}$ is slightly larger than the theoretically predicted one. This may be attributed to the presence of a small orbital moment in addition to the spin moment. An orbital moment must be present because of the observed anisotropy of the *g*-factor [14].

For spins without coupling one expects an increase of magnetic susceptibility with decreasing temperature according to a Curie law. The experimental curve in contrast decreases upon cooling, indicating antiferromagnetic coupling. There is a pronounced maximum at $T_N$=40 K that is evident from Fig. 2 c. We associate this maximum with a low-temperature phase with weak antiferromagnetic interaction. The estimation of the number of unpaired spins at room temperature from the magnetic susceptibility using the formula $S(S+1)=3k_BT\chi/(N_Ag^2\mu_B^2)$ yields a spin value $S=0.525$ for one DOEO dimer. This value is 5% larger than the expected value $S=\frac{1}{2}$. The deviation may be attributed to the orbital magnetic moment as discussed above.

The effective magnetic moment was calculated by the formula $\mu_{eff}=(3\chi_{dc}k_BT/N_A\mu_B)^{1/2}\approx(8MT/vH)^{1/2}$, where $v$ is the amount of material in mol, $k_B$ is the Boltzmann constant, $T$ is the temperature, $N_A$ is the Avogadro constant, and $\mu_B$ is the Bohr magneton. The effective magnetic moment is a rough measure for the size of the spin at each site, neglecting coupling and impurities. The effective moment determined here is close to the expected value $\mu_{eff}=\mu_Bg(S(S+1))^{1/2}=1.73\ \mu_B$ for independent centers with spin $S=\frac{1}{2}$. The experimental value varies between 1 $\mu_B$ (at 50 K) and 2.5 $\mu_B$ (at 300 K) and is equal to the theoretical value for $S=\frac{1}{2}$ at $T\approx150$ K.

The important feature from our point of view is the agreement in the temperature dependences of the three main characteristics of charge carriers − resistivity [11] (Fig. 2 a), integral intensity obtained from the ESR measurements [14] (Fig. 2 b) and magnetic susceptibility from the SQUID data (Fig. 2 c). The critical temperature $T$=70 K shows up as a minimum in resistance, in the appearance of the additional line in the integral ESR intensity and a minimum in susceptibility. The low-temperature phase is insulating according to a steep increase in resistivity. Charge carriers in this phase are localized in two types of centers according to two ESR lines with different symmetry of the crystal field [14]. Moreover, it is characterized by an antiferromagnetic coupling of spins. Similarly to (BEDT-TTF)(TCNQ) [18] we find an *antiferromagnetic insulator* at low temperatures.

The next temperature region 70–120 K shows the characteristics of a *metallic phase* with almost linear increase of resistivity and magnetic susceptibility with increasing temperature. Due to the specific ESR technique that usually detects localized electrons (or holes), it is impossible to observe changes in itinerant electrons [15]. There are no changes and jumps in the temperature dependence of the integral ESR intensity of the crystal (Fig. 2 b). With increasing temperature the ESR intensity rises in the metallic regime because more and more electrons lose their itinerant character.

At the transition point to the high-temperature phase at $T>120$ K the resistivity of $(DOEO)_4[HgBr_4]\cdot TCE$ suddenly changes its slope and drops. It shows similar behavior to well-known $\kappa$-$(BEDT-TTF)_2Cu[N(CN)_2]Cl$ [17] where with increasing temperature hopping conductivity appears and it does not allow to show a metallic-like behavior anymore. This is the *bad metal (anomalous metal)* phase. In this phase the magnetic susceptibility is nearly constant.

The proposed phase diagram for the charge-transfer salt $(DOEO)_4[HgBr_4]\cdot TCE$ (Fig. 2 d) is compatible with the conceptual phase diagram by K. Miyagawa et. al. for the ET-family [19]. It seems that at low temperatures it is very close to the boundary between AF insulator and superconductor. With increasing temperature a transition to a metal and finally to a bad metal is observed.

**Conclusions**

We studied the organic charge transfer salt $(DOEO)_4[HgBr_4]\cdot TCE$. The comparison of resistivity, ESR and SQUID measurements reveals a phase model for DOEO charge transfer salts. At low temperatures we find an antiferromagnetic insulator that transforms into a paramagnetic insulator at the Néel temperature of 40 K. At 70 K a phase transition to a metallic phase occurs. With increasing temperature up to 120 K the metallic behavior changes to a bad metal due to thermally activated molecular movement, loss of itinerant character and increase of hopping conductivity.

**Acknowledgment**

We thank the Deutsche Forschungsgemeinschaft (SFB/TR 49) for financial support.